\documentclass[aps,prd,amsmath]{revtex4}
\usepackage{times,amsbsy,amsfonts,graphicx,float}
\usepackage{color,morefloats,rotating,srcltx,slashed}
\usepackage{multirow,bm,verbatim,tabularx,bbding,threeparttable}
\definecolor{dblue}{rgb}{0.00,0.00,0.75}
\usepackage[colorlinks,urlcolor=dblue,linkcolor=dblue,citecolor=dblue]{hyperref} 
\usepackage{setspace}
\allowdisplaybreaks[4]

\def \babar {{\it BABAR} }

\begin{document} 

\title{Traces of the new $a_0(1780)$ resonance in the $J/\psi\to \phi K^+K^-(K^0\bar{K}^0)$ reaction} 
                    
\author{Luciano M. Abreu$^{1,2}$}     \email{luciano.abreu@ufba.br}
\author{Wen-Fei Wang$^{2,3}$}     \email{wfwang@ific.uv.es}
\author{Eulogio Oset$^2$}             \email{oset@ific.uv.es}
		
\affiliation{$^1$Instituto de F\'{\i}sica, Universidade Federal da Bahia, Campus Universit\'{a}rio de Ondina, 40170-115 Bahia, Brazil\\
                 $^2$Departamento de F\'{\i}sica Te\'orica and IFIC, Centro Mixto Universidad de Valencia-CSIC, \\
	                 Institutos de Investigaci\'on de Paterna, Aptdo.~22085, 46071 Valencia, Spain\\           
	         $^3$Institute of Theoretical Physics, Shanxi University, Taiyuan, Shanxi 030006, China
}	

\date{\today}

\begin{abstract}
We study the $J/\psi \to \phi K \bar K$ decay, looking for differences in the production rates of $K^+ K^-$ or $K^0 \bar K^0$ in the region of 1700-1800 MeV, where two resonances appear dynamically generated from the vector-vector interaction. Two resonances are known experimentally in that region, the $f_0(1710)$ and a new resonance reported by the BABAR and BESIII collaborations. The $K \bar K$ should be produced with $I=0$ in that reaction, but due to the different $K^{*0}$ and $K^{*+}$ masses some isospin violation appears. Yet, due to the large width of the $K^*$, the violation obtained is very small and the rates of $K^+ K^-$ or $K^0 \bar K^0$ production are equal within $5\%$. However, we also find that due to the step needed to convert two vectors into $K \bar K$, a shape can appear in the $K \bar K$ mass distribution that can mimic the $a_0$ production around the $K^* \bar K^*$ threshold, and is simply a threshold effect.
\end{abstract}

\maketitle

\section{Introduction}                  
At a time where it looked like hadron spectroscopy in the light quark sector at energies below $2$ GeV was more or less settled, 
the discovery of a new meson resonance with isospin $I=1$ around $1700$-$1800$ MeV has come as a real surprise.
The saga began with the \babar experiment studying the $\pi^+\eta, \pi^-\eta$ mass distributions from the 
$\eta_c \to\pi^+\pi^-\eta$ decay, from where some peaks were observed around $1700$ MeV~\cite{babar}.
The search continued at BESIII, where in Ref.~\cite{bes1} a branching fraction
\begin{eqnarray}
 \mathcal{B}(D^+_s\to\pi^+ ``f_0(1710)"; ``f_0(1710)"\to K^+K^-)=(1.0\pm0.2\pm0.3)\times10^{-3}
 \label{eq1}
\end{eqnarray}
was found, while in Ref.~\cite{bes2} a related branching fraction was reported as
\begin{eqnarray}
 \mathcal{B}(D^+_s\to\pi^+ ``f_0(1710)"; ``f_0(1710)"\to K^0_SK^0_S)=(3.1\pm0.3\pm0.1)\times10^{-3}
 \label{eq2}
\end{eqnarray}
We write $``f_0(1710)"$ because this was the resonance looked for in the experiments, which is well known and reported in the 
PDG~\cite{pdg}. However, the combined information of Eqs.~(\ref{eq1})-(\ref{eq2}) is most surprising since from there one finds 
the ratio
\begin{eqnarray}
 R_1=\frac{\Gamma(D^+_s\to \pi^+K^0\bar{K}^0)}{\Gamma(D^+_s\to \pi^+K^+K^-)}=6.20\pm0.67,
 \label{eq_ratio}
\end{eqnarray}
while this ratio should be $1$ if only an $f_0(1710)$ with $I=0$ existed. The large deviation of Eq.~(\ref{eq_ratio}) from unity implied
the existence of a structure with $I=1$. While both states with $I=0$ or $I=1$ would provide individually the same strength for 
$K^+K^-$ or $K^0\bar{K}^0$ production, the simultaneous existence of the $I=0$ $f_0(1710)$ and another $I=1$ structure 
would lead to different rates through the interference of the two structures. The new $I=1$ state is reported in the region of 
$1710$ MeV where the $f_0(1710)$ appears.

Interestingly, there were theoretical prediction for this $I=1$ state. Indeed, in the relativized quark model of Godfrey and 
Isgur~\cite{goisgur}, the $f_0(1710)$ appears as an $I^G(J^{PC})=0^+(0^{++})$ state from the $2^3P_1$ quark antiquark 
configuration. But an $I=1$ state around the same energy with the same configuration is reported there. In Ref.~\cite{entem}
similar findings are reported, while in the related work of Ref.~\cite{vijande} and $I=0$ state is obtained with the same configuration 
but no mention is made of an $I=1$ state. The models mentioned above considered only the excitations of $u, d$ quarks and, 
concerning the $I=1$ state, in the $q\bar{q}$ picture an $a^+_0$ has a $u\bar{d}$ structure, but the $f_0(1710)$ decays mostly in 
$K\bar K$, $\eta\eta$ with only $4\%$ branching fraction to $\pi\pi$ decay, suggesting that that state should have a large 
component of $s\bar s$ quarks. The observation of the new $I=1$ structure in $K\bar K$ in~\cite{bes1,bes2} also points in the 
same direction.

A different picture of these states is offered in the work of~\cite{geng} where the chiral unitary approach used to study the 
interaction of pseudoscalar mesons, from where the low lying scalar mesons $f_0(500)$, $f_0(980)$, $a_0(980)$ and 
$K^*_0(700)$ emerge, is extrapolated to study the interaction of vector mesons among themselves. The local hidden 
gauge approach~\cite{hidden1,hidden2,hidden4,hideko}, shown to lead to the chiral Lagrangians in the pseudoscalar 
sector~\cite{derafael}, provides also the interaction between vector mesons and is used as a source of the vector meson potential 
in Ref.~\cite{geng} which, via the Bethe-Salpeter equation, implements unitarity in coupled channels. Among many 
other states of spin $J=0,1,2$, two states appeared around the $1700$-$1800$ MeV region, an $I=0$ state at $1721$ MeV 
and width $133$ MeV, which was associated to the $f_0(1710)$, and an $I=1$ state at $1777$ MeV with 
$\Gamma=148$ MeV (we shall call it $a_0(1780)$) for which there was no counterpart in the PDG at the time of the 
prediction. More recently similar results concerning these two states are also found in~\cite{guomeiss} using dispersion relations 
in the region of validity of the model (see Ref.~\cite{careful} for further discussions). The $f_0(1710)$ couples to the channels 
$K^*\bar K^*, \rho\rho, \omega\omega, \omega\phi$ and $\phi\phi$, while the $a_0(1780)$ couples to $K^*\bar K^*, \rho\omega,
\rho\phi$, with the $K^*\bar K^*$ channel being dominant for the two states. Given the fact that in~\cite{geng} the two 
$I=0,I=1$ states were obtained, it was a challenge to see if the results of BESIII could be reconciled within that picture, 
a task that was undertaken in Ref.~\cite{daigeng}, where, considering the weak decay mechanisms of the $D^+_s$ from external 
and internal emission~\cite{chau}, the ratio $R_1$ of Eq.~(\ref{eq_ratio}) could be reproduced. In Ref.~\cite{daigeng} using 
the same mechanism suited to reproduce $R_1$ of Eq.~(\ref{eq_ratio}), predictions were made for the branching fraction 
of the decay of $D^+_s\to \pi^0a_0(1780); a_0(1780)\to K^+K^0_S$, which resulted in a fair agreement with the measurements 
done later at BESIII in Ref.~\cite{besafter}, where the $a_0$ state was reported at $1817$ MeV with mass and width 
\begin{eqnarray}
     M_{a_0}&=&1817\pm 8_{\rm stat.} \pm 20_{\rm sys.}~{\rm MeV},\nonumber\\
   \Gamma_{a_0}&=&97\pm22_{\rm stat.} \pm 15_{\rm sys.}~{\rm MeV}.
   \label{data_a0}
\end{eqnarray}
With uncertainties in the position of the $a_0$ resonance, it is quite fair to assume that all experiments~\cite{babar,bes1,bes2,
besafter} are seeing the same state, a new $a_0$ resonance with mass $1700$-$1800$ MeV, and this has been assumed 
in theoretical works triggered by these findings. In~\cite{wangzou}, where the idea of~\cite{geng} is retaken, it was shown 
that the addition of pseudoscalar channels to the vector-vector coupled channels of~\cite{geng} does not significantly change 
the results of~\cite{geng}. Yet, some discussion is done on how the appearance of the new $a_0$ resonance has some influence 
on the ongoing discussions about the $f_0(1500)$ or $f_0(1710)$ resonances to correspond to glueball states~\cite{amsler}.  
In Ref.~\cite{shilin} the discovery of a new $a_0$ resonance is used to reclassify states in Regge trajectories, suggesting 
that a new $a_0$ resonance should appear at $2115$ MeV.

The work of~\cite{daigeng} has been further improved in~\cite{gengxie} including the $D^+_s\to K^{*+}\bar K^0\to \pi^+K^0_SK^0_S$
mechanism and in~\cite{wangeng} including the extra $D^+_s\to \pi^0a_0(980)\to \pi^0K^+K^0_S$ mechanism. With this 
extended model over the $\pi^+$-vector-vector ($\pi^+$VV) decay channels considered in~\cite{daigeng}, a good reproduction 
of the different invariant mass distributions is obtained. A perspective view of the repercussions of the new found resonance is given in~\cite{Oset:2023hyt}.

In the present work, we look into a different reaction exploiting the strong interaction and looking at the 
$J/\psi\to \phi K^+K^-(K^0\bar K^0)$ decays. Since both $J/\psi$ and $\phi$ have $I=0$, the $K\bar K$ system should be created 
in $I=0$ and hence one should have the same rate of production for $K^+K^-$ and $K^0\bar K^0$. However, the fact that $K^{*+}$ 
and $K^{*0}$ have different masses will induce some isospin violation~\cite{achasov} and one could find traces of the 
$a_0(1780)$ in the reaction. The purpose of this work is to investigate in detail this possibility.

\section{Formalism: effective interactions and $VV$ states}\label{sec:2}           

\begin{figure}[tbp]   
  \centering
  \includegraphics[width=7cm]{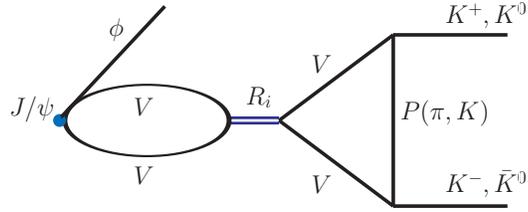}
  \caption{Mechanism for the production of the resonances $R_i=f_0(1710), a_0(1780)$ with further decay to 
                $K^+K^-, K^0\bar K^0$; $P$ denotes a pseudoscalar exchange.
              }
  \label{fig1}
\end{figure}

Since the $f_0(1710)$ and $a_0(1780)$ resonances are obtained from the interaction of vector-vector channels in~\cite{geng}, 
we have to let the $J/\psi$ decay into $\phi VV$ and then let the $VV$ pair interact to produce the resonances. Finally the resonances 
must decay into $K^+K^-$ or $K^0\bar K^0$. The whole process is depicted in Fig.~\ref{fig1}.

The first step is to produce the $\phi$ and the different channels that produce the $f_0, a_0$ resonances. For this we have to 
look at the decay of $J/\psi$ into $VVV$, selecting the terms that contain at least one $\phi$ field. The idea is to exploit the 
fact that the $J/\psi(c\bar c)$ state is a singlet of SU$(3)$ (SU$(3)$ scalar), hence the combination of $VVV$ has to be a SU$(3)$ 
scalar. This idea has been used in related problems like $\chi_{c1}\to \eta\pi^+\pi^-$~\cite{liangxie}, $\eta_c\to \eta\pi^+\pi^-$
\cite{debastiani}, $\chi_{cJ}\to \phi K^*K$~\cite{sakailiang}, $J/\psi\to\gamma\pi\pi,\gamma\pi^0\eta$~\cite{liangtoledo}, 
$J/\psi\to\phi(\omega)VV$~\cite{raqueldai} and $\chi_{c1}\to PVV$~\cite{diasliang}. Concretely, if we look at the $q_iq_j$ matrix 
in SU$(3)$ ($i,j=u,d,s$) and write it in terms of physical mesons, we have the following representations for pseudoscalars and 
vectors, using the standard mixing of $\eta, \eta^\prime$~\cite{bramon}
\begin{equation}
P = \left(
           \begin{array}{ccc}
             \frac{1}{\sqrt{2}}\pi^0 + \frac{1}{\sqrt{3}}\eta + \frac{1}{\sqrt{6}}\eta' & \pi^+ & K^+ \\
             \pi^- & -\frac{1}{\sqrt{2}}\pi^0 + \frac{1}{\sqrt{3}}\eta + \frac{1}{\sqrt{6}}\eta' & K^0 \\
            K^- & \bar{K}^0 & -\frac{1}{\sqrt{3}}\eta + \sqrt{\frac{2}{3}}\eta' \\
           \end{array}
         \right),
  \label{eq3}
\end{equation}

\begin{equation}
V = \left(
           \begin{array}{ccc}
             \frac{1}{\sqrt{2}}\rho^0 + \frac{1}{\sqrt{2}}\omega  & \rho^+ & K^{*+} \\
             \rho^- & -\frac{1}{\sqrt{2}}\rho^0 + \frac{1}{\sqrt{2}}\omega  & K^{*0} \\
            K^{*-} & \bar{K}^{*0} & \phi \\
           \end{array}
         \right).
   \label{eq4}
\end{equation}

We can have three SU$(3)$ invariants made from the $V$ matrices 
\begin{equation}
 \langle VVV\rangle, \quad \langle VV\rangle \langle V\rangle, \quad  
         \langle V\rangle\langle V\rangle\langle V\rangle,
\end{equation}
with $\langle \cdot\cdot\cdot\rangle$ indicating the trace in SU$(3)$. Then we keep the terms having at least one $\phi$ field and 
we find the following combinations 
\begin{eqnarray}
  &&1)~\langle VVV\rangle: ~3(K^{*+}K^{*-}+K^{*0}\bar K^{*0})\phi+\phi\phi\phi, \nonumber \\
  &&2)~ \langle VV\rangle \langle V\rangle: ~\sqrt2\omega\phi\phi+(\rho^0\rho^0+\rho^+\rho^-+\rho^-\rho^++
                   \omega\omega+2K^{*+}K^{*-}+2K^{*0}\bar K^{*0})\phi +\phi\phi\phi,\qquad       \nonumber\\
  &&3)~ \langle V\rangle\langle V\rangle\langle V\rangle: ~4\omega\omega\phi+2\sqrt2\omega\phi\phi+\phi\phi\phi.
  \label{eq6}             
\end{eqnarray}
Yet, in the works mentioned above it was always the trace of the product of three matrices the one that dominates, in  our case 
$\langle VVV\rangle$. Actually, we can be more precise taking advantage of the work done in~\cite{raqueldai}. In this last work 
the same problem $J/\psi\to \phi(\omega)VV$ was studied, however, with a different aim. Indeed, the purpose of 
Ref.~\cite{raqueldai} was to correlate the ratios of production of different resonances which are generated from the $VV$ 
interactions according to~\cite{geng}. In this sense several ratios could be calculated theoretically relating the rates of
$J/\psi\to\phi f_2(1270)$, $J/\psi\to \phi f^\prime_2(1525)$, $J/\psi\to\omega f_2(1270)$, $J/\psi\to \omega f^\prime_2(1525)$
and $J/\psi\to K^{*0}\bar K^{*0}_2(1430)$ on one side and $J/\psi\to \omega f_0(1320)$, $J/\psi\to \phi f_0(1320)$, 
$J/\psi\to \omega f_0(1710)$, $J/\psi\to \phi f_0(1710)$ on the other side.  A good agreement was found with four known 
experimental ratios. The study, in which weights $\alpha$ and $\beta$ were given to the $\langle VVV\rangle$ and 
$\langle VV\rangle \langle V\rangle$ structures (the $\langle V\rangle^3$ structure is disfavored), concluded that the 
$\langle VVV\rangle$ one was favored and the best fit to the data was obtained with the value $\beta/\alpha=0.32$.
We will use this finding here and take the same two structures with the same ratio between them.

The $VV$ states that we have are given by
\begin{eqnarray}
&&|K^*\bar K^*, I=0\rangle = \frac{-1}{\sqrt2}(K^{*+}K^{*-}+K^{*0}\bar K^{*0}),                          \nonumber\\
&&|K^*\bar K^*, I=1, I_3=0\rangle = \frac{1}{\sqrt2}(K^{*0}\bar K^{*0}-K^{*+}K^{*-}),                  \nonumber\\
&&|\rho\rho, I=0, I_3=0\rangle = \frac{-1}{\sqrt6}(\rho^{+}\rho^{-}+\rho^{-}\rho^{+}+\rho^{0}\rho^{0}),                  \nonumber\\
&&|\omega\omega\rangle\to \frac{1}{\sqrt2}|\omega\omega\rangle,                  \nonumber\\
&&|\phi\phi\rangle\to \frac{1}{\sqrt2}|\phi\phi\rangle,                  \nonumber\\
&&|\rho^0\omega\rangle\to |\rho^0\omega\rangle,                  \nonumber\\
&&|\rho^0\phi\rangle\to |\rho^0\phi\rangle,    
\label{eqwfun} 
\end{eqnarray}
where the phases of the isospin multiplets $(K^{*+}, K^{*0})$, $(\bar K^{*0}, -K^{*-})$, $(-\rho^+,\rho^0,\rho^-)$ are taken, and the 
$\frac{1}{\sqrt2}$ extra factor is implemented in the identical particles for practical reasons in the counting of states in the 
intermediate loops (unitary normalization). It is not surprising that all the structures in $1), 2)$ and $3)$ of Eq.~(\ref{eq6}) filter the 
$I=0$ states since $\phi$ and $J/\psi$ have both $I=0$.

\section{Scattering matrices in $I=0, I=1$}\label{sec:3}
The next step done in~\cite{geng} is to evaluate the transition potentials between the states in $I=0$ and $I=1$ independently 
$V_{ij}$, and then construct the scattering matrix $T_{ij}$ using the Bethe-Salpeter equation in its factorized form
\begin{eqnarray}
      T=[1-VG]^{-1}V,  \label{eq7}
\end{eqnarray} 
where $G$ is the loop function for two vectors intermediate states
\begin{eqnarray}
      G={\rm diag}[G_l(\sqrt s)],  \label{eq8}
\end{eqnarray} 
and $G_l$ is regularized with dimensional or cut-off regularization, with
\begin{eqnarray}
G_{l}(\sqrt{s})&=& \frac{1}{16 \pi^2} \left\{ a(\mu) + \ln\frac{M_l^2}{\mu^2} + \frac{m_l^2-M_l^2 + s}{2s} 
                             \ln \frac{m_l^2}{M_l^2} \right. \nonumber\\ 
& & \phantom{\frac{1}{16\pi^2}} + \frac{q_l}{\sqrt{s}} \left[ \ln(s-(M_l^2-m_l^2)+2q_l\sqrt{s})
                                                     + \ln(s+(M_l^2-m_l^2)+2 q_l\sqrt{s}) \right. \nonumber  \\
& & \left. \phantom{\frac{1}{16 \pi^2} + \frac{q_l}{\sqrt{s}}} \left. \hspace*{-0.3cm} - \ln(-s+(M_l^2-m_l^2)+2 q_l\sqrt{s})
                                                           -\ln(-s-(M_l^2-m_l^2)+2 q_l\sqrt{s}) \right] \right\},
\label{eq9}
\end{eqnarray}
in the dimensional regularization with $\mu$ a scale taken $\mu=1000$ MeV in~\cite{geng}, $M_l, m_l$ the masses of the 
intermediate vectors and $a(\mu)$ a subtraction constant, usually fitted to some data. The value $a=-1.726$ was used in 
\cite{geng} and so we shall do here. The value of $q_l$ is the on-shell momentum of the $VV$ channel for energy $\sqrt s$. 
For the cut-off regularization we have
\begin{equation}
   G(\sqrt s)=\int\limits_{|\vec{q}|\le q_\mathrm{max}}\frac{d^3q}{(2\pi)^3}\frac{\omega_l+\omega_l^\prime}
                  {2\omega_l\omega_l^\prime[(P^{0})^2-(\omega_l+\omega_l^\prime)^2+i\epsilon]},
   \label{eq10G}          
\end{equation}
with $\omega_l=\sqrt{\vec q^{\,2}+m^2_l}$ and $\omega^\prime_l=\sqrt{\vec q^{\,2}+m^{\prime2}_l}$, 
the $q_\mathrm{max}$ indicates the range of the $V$ interaction in momentum space~\cite{danijuan,daisong}, and is also
tuned to some experimental data. The value of $q_{max}$ used in~\cite{geng} was $1000$ MeV that we also use here.

The purpose here is to see how isospin is broken due to the different masses of the $K^{*+}$ and $K^{*0}$, since this makes the 
$G$ functions different for $K^{*+}K^{*-}$ or $K^{*0}\bar K^{*0}$, and this will modify both the scattering matrices and the 
mechanism of Fig.~\ref{fig1}. To implement these changes in the $T$ matrix of Eq.~(\ref{eq7}) we would have to go back to the 
formalism of Ref.~\cite{geng} and redo all the calculations with the new loops keeping the physical masses of the charged and 
neutral $K^*$ (average values were used in~\cite{geng} to keep isospin symmetry). Yet, we can circumvent this fact with the 
following procedure: we take the $I=0$ and $I=1$ $T$ matrices from~\cite{geng} using the pole approximation in a Breit-Wigner 
(BW) form
\begin{eqnarray}
    T^{(I=0)}_{ij}=\frac{g_ig_j}{M^2_{\rm inv}-M^2_{f_0}+iM_{\rm inv}\Gamma_{f_0}},\quad
     T^{(I=1)}_{ij}=\frac{g_ig_j}{M^2_{\rm inv}-M^2_{a_0}+iM_{\rm inv}\Gamma_{a_0}},
     \label{eq10}
\end{eqnarray}
with the coupling $g_i$ of the resonances obtained in Ref.~\cite{geng}, given in Tables~\ref{table1}, \ref{table2} and 
\begin{eqnarray}
   && M_{f_0}=1721~{\rm MeV}, \quad \Gamma_{f_0}=133~{\rm MeV}, \nonumber\\
   && M_{a_0}=1777~{\rm MeV}, \quad \Gamma_{a_0}=148~{\rm MeV}.
   \label{eq11}
\end{eqnarray}
We should mention that $\Gamma_{f_0}, \Gamma_{a_0}$ contain the decay of the resonances into two pseudoscalars, 
which was evaluated in~\cite{geng} via box diagrams with $PP$ in the intermediate states.

\begin{table*}[!]
\centering
\caption{Couplings $g_i$ of the $f_0(1721)$ into the different channels from Ref.~\cite{geng}.
             }
\label{table1}
\setlength{\tabcolsep}{8pt}
\setstretch{1.2}
\begin{tabular}{lccccc}
\hline 
\hline
       &  $K^*\bar{K}^*$ & $\rho\rho$ & $\omega\omega$  & $\omega\phi$  & $\phi\phi$ \\
 $g_i$ &  $7124+i96$ & $-1030+i1086$ & $-1763+i108$ & $3010-i210$ & $-2493-i204$ \\
\hline
\hline
\end{tabular}
\end{table*}

\begin{table*}[!]
\centering
\caption{Same as in Table~\ref{table1} but for $a_0(1780)$.
             }
\label{table2}
\setlength{\tabcolsep}{8pt}
\setstretch{1.2}
\begin{tabular}{lccc}
\hline 
\hline
       &  $K^*\bar{K}^*$ &  $\rho\omega$  & $\rho\phi$ \\ 
 $g_i$ &  $7525-i1529$ &  $-4042+i1391$ & $4998-i1872$  \\  
\hline
\hline
\end{tabular}
\end{table*}

Next, in order to mix $I=0$ and $I=1$ we write now the channels $K^{*+}K^{*-}(1)$, $K^{*0}\bar K^{*0}(2)$, $\rho\rho(I=0)(3)$,
$\omega\omega(4)$, $\omega\phi(5)$, $\phi\phi(6)$, $\rho\omega(7)$ and $\rho\phi(8)$. Taking the wave functions of 
Eq.~(\ref{eqwfun}) we will have in the isospin conserving scheme
\begin{eqnarray}
  T_{K^{*+}K^{*-},K^{*+}K^{*-}} &=& \frac12 (T^{I=0}_{K^*\bar{K}^*,K^*\bar{K}^*} + T^{I=1}_{K^*\bar{K}^*,K^*\bar{K}^*}), \nonumber\\ 
  T_{K^{*0}\bar K^{*0},K^{*0}\bar K^{*0}} &=& \frac12 
                        (T^{I=0}_{K^*\bar{K}^*,K^*\bar{K}^*} + T^{I=1}_{K^*\bar{K}^*,K^*\bar{K}^*}), \nonumber\\   
  T_{K^{*+}K^{*-},K^{*0}\bar K^{*0}} &=& \frac12 
                         (T^{I=0}_{K^*\bar{K}^*,K^*\bar{K}^*} - T^{I=1}_{K^*\bar{K}^*,K^*\bar{K}^*}), \nonumber\\          
    T_{K^{*+}K^{*-},i}    &=&   -\frac{1}{\sqrt2} T^{I=0}_{K^*\bar{K}^*,i}; ~i=3,4,5,6, \nonumber\\          
    T_{K^{*0}\bar K^{*0},i}    &=&   -\frac{1}{\sqrt2} T^{I=0}_{K^*\bar{K}^*,i}; ~i=3,4,5,6, \nonumber\\    
    T_{K^{*+}K^{*-},j}    &=&   -\frac{1}{\sqrt2} T^{I=1}_{K^*\bar{K}^*,j}; ~j=7,8, \nonumber\\          
    T_{K^{*0}\bar K^{*0},j}    &=&   \frac{1}{\sqrt2} T^{I=1}_{K^*\bar{K}^*,j}; ~j=7,8.   
    \label{eq12}                         
\end{eqnarray}

Eq.~(\ref{eq12}) corresponds to 
\begin{eqnarray}
      T=[1-VG]^{-1}V,  \label{eq13}
\end{eqnarray} 
in the isospin symmetric case with $G$ the diagonal matrix $G={\rm diag}[G_l]$, $l=1$-$8$. We can obtain $V$ from Eq.~(\ref{eq13}) 
via
\begin{eqnarray}
   T^{-1}=V^{-1}-G; \quad V^{-1}=T^{-1}+G,   \label{eq14}
\end{eqnarray}
and then we can construct the new $T$ matrix ($\tilde{T}$) with isospin violation from different $K^{*+}K^{*-}$ and 
$K^{*0}\bar K^{*0}$ loops using the same $V$ matrix as
\begin{eqnarray}
   \tilde{T}^{-1}=V^{-1}-\tilde{G}=T^{-1}+G-\tilde{G},   \label{eq15}
\end{eqnarray}
with $G$ obtained with averaged $K^{*+}$, $K^{*0}$ matrices and $\tilde{G}$ calculated with the $K^{*+}K^{*-}$ and 
$K^{*0}\bar K^{*0}$ masses. Hence 
\begin{eqnarray}
   \tilde{T}^{-1}=T^{-1}+\delta G_{ij},   \label{eq16}
\end{eqnarray}
with
\begin{eqnarray}
   \delta G_{K^{*+}K^{*-}}&=&G_{K^{*}\bar K^{*}}-\tilde{G}_{K^{*+}K^{*-}}, \nonumber\\
   \delta G_{K^{*0}\bar K^{*0}}&=&G_{K^{*}\bar K^{*}}-\tilde{G}_{K^{*0}\bar K^{*0}}, \nonumber\\  
   \delta G_{j} &=& 0, \quad j=3,4,5,6,7,8.  \label{eq17}
\end{eqnarray}
Then we invert $\tilde{T}^{-1}$ and we already have the isospin violating $T$ matrix which mixes $I=0$ and $I=1$. There is 
a caveat, however, about using this procedure. The matrix $T_{ij}$ constructed from Eq.~(\ref{eq10}) is not invertible since 
it has determinant zero.
But this is no problem for the use of the proposed scheme. We resort to a trick which is to multiply the diagonal terms $T_{ii}$ 
by a factor close to $1$ ($1.005$ in practice for our case) and then $T^{mod}_{ij}$ is already invertible. All we have to check is 
that inverting $({T^{-1}})^{mod}_{ij}$ we get the original $T$ matrix with great precision and no artificial numerical results are 
introduced. We have checked that this is the case and then, since relative changes of $\delta G$ are much bigger than $1$ in 
$10^5$, the resulting $\tilde T$ matrix is meaningful. An alternative technical way is to invert Eq.~(\ref{eq15}) which gives 
\begin{eqnarray}
   \tilde{T}=(1+T \, \delta G)^{-1} \,T,   \label{eq15prime}
\end{eqnarray}
which overcomes the problem of the inversion of $T$. 

\section{$J/\psi\to \phi K^+K^-(K^0\bar{K}^0)$ production mechanism}

The mechanism for production of final $K^{*+}K^{*-}$, $K^{*0}\bar K^{*0}$ is shown in Fig.~\ref{fig1}. We shall take all 
$VV$ channels that contribute and we must evaluate the triangle loop. Denoting $V_{VV,j}$ the triangle loops, with 
$j=K^{*+}K^{*-}, K^{*0}\bar K^{*0}$, we first show that only the $VV=K^{*+}K^{*-}, K^{*0}\bar K^{*0}$ intermediate states are 
relevant in the loop. Indeed, the $K^{*}\bar K^{*}$ channels are there with stronger coupling to both the $f_0$ and $a_0$ resonances. 
Second, for $VV=K^{*+}K^{*-}, K^{*0}\bar K^{*0}$ the pseudoscalar $P$ meson exchanged is a pion, while for the other
$VV$ channels the meson exchanged is a kaon, and due to the larger mass of the kaon the propagator is reduced versus 
the one of the pion. Due to this, the $K^{*}\bar K^{*}$ channels are the relevant ones and we only keep them in the triangle 
loop. The fact that we are interested in the isospin breaking due to different $K^{*+}, K^{*0}$ masses further stresses the 
special role of the $K^{*}\bar K^{*}$ in the triangle loop. With these considerations we can already write the amplitudes for the 
$J/\psi\to \phi K^+K^-(K^0\bar{K}^0)$ production through the mechanism of Fig.~\ref{fig1}
\begin{eqnarray}
 t_i(M_{inv}) = \sum_j W_j G_j(M_{inv})(\tilde T_{j, K^{*+}K^{*-}} V_{K^{*+}K^{*-},i}  
                                +\tilde T_{j, K^{*0}\bar K^{*0}} V_{K^{*0}\bar K^{*0},i} ),
            \label{eq18}
\end{eqnarray}
with $j=K^{*+}K^{*-}, K^{*0}\bar K^{*0}, \omega\omega, \omega\phi, \phi\phi, \rho\rho$ and $i=K^{+}K^{-}, K^{0}\bar K^{0}$,
$W_j$ are the weights by which the different $j$ channels (only those of $I=0$) are produced according to the expressions 
in Eq.~(\ref{eq6}). We shall give weight $\alpha, \beta, \gamma$ to the $\langle VVV\rangle$, $\langle VV\rangle \langle V\rangle$ 
and  $\langle V\rangle^3$ structures, respectively, although at the end we shall take $\gamma=0$, $\beta/\alpha=0.32$ as found 
in Ref.~\cite{raqueldai}. Taking into account the symmetry factor $n!$ for $n$ identical particles and the normalization used in 
Eq.~(\ref{eqwfun}), we find the weights
\begin{eqnarray}
   &&W_{K^{*+}K^{*-}}=3(\alpha+\beta); \qquad W_{K^{*0}\bar K^{*0}}=3(\alpha+\beta); \quad\;\;\;
        W_{\omega\omega}=\sqrt2(\beta+4\gamma); \nonumber\\
   &&W_{\omega\phi}=2(\sqrt2\beta+2\sqrt2\gamma); \quad\; W_{\phi\phi}=\frac{6}{\sqrt 2}(\alpha+ \beta+ \gamma); \quad
        W_{\rho\rho}=-\sqrt{\frac32}\beta.
     \label{eq19}
\end{eqnarray}

\begin{figure}[tbp]   
  \centering
  \includegraphics[width=8cm]{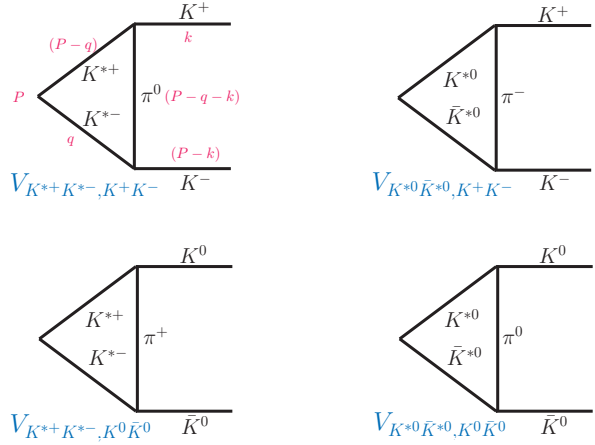}
  \caption{Triangle loop functions for the evaluation of $V_{K^{*}\bar K^{*}, K\bar K}$
              }
  \label{fig2}
\end{figure}

The evaluation of the vertex requires the use of the $VPP$ Lagrangian 
\begin{align}
    \mathcal{L}_{\mathrm{VPP}} &= -i g\left\langle\left[P, \partial_{\mu} P\right] V^{\mu}\right\rangle,    
     \label{eq20}
\end{align}
with the $P,V$ given by Eqs.~(\ref{eq3})-(\ref{eq4}), the coupling  $g=\frac{m_V}{2f_\pi}$ with $m_V=800$ MeV and the pion 
decay constant $f_\pi=93$ MeV.  We plot in Fig.~\ref{fig2} the triangle diagrams showing explicitly the momenta.

The triangle loop function is given by 
\begin{eqnarray}
 -iV_j=-i\int \frac{d^4q}{(2\pi)^4}\frac{i}{q^2-M^2_{K^*_j}+i\epsilon} \frac{i}{(P-q)^2-M^2_{K^*_j}+i\epsilon }
                     \frac{i}{(P-q-k)^2-M^2_{\pi}+i\epsilon } \tilde W_j (2{\vec k+\vec q})^2 g^2, 
    \label{eq21}
\end{eqnarray}
with the weighs
\begin{equation}
   \tilde W_j \equiv  \left\{ 
         \begin{aligned}
           & -\frac{1}{2}, \quad {\rm for}\quad  K^{*+}K^{*-}, K^{+}K^{-},             \\
           & -1,                \quad\; {\rm for}\quad  K^{*0}\bar K^{*0}, K^{+}K^{-},                     \\
           & -1,                \quad\; {\rm for}\quad  K^{*+}K^{*-}, K^{0}\bar K^{0},                \\
           & -\frac{1}{2}, \quad  {\rm for}\quad  K^{*0}\bar K^{*0}, K^{0}\bar K^{0},          
         \end{aligned} \right.
     \label{eq22}
\end{equation}
The integration is made easier by taking only the positive energy part of the propagators for the $K^*$ states, given the large 
mass of $K^*$, while keeping both terms for the pion propagator, with the separation of positive energy parts as
\begin{eqnarray}
 \frac{1}{q^2-m^2+i\epsilon}=\frac{1}{2\omega(\vec q)}\left(\frac{1}{q^0-\omega(\vec q)+i\epsilon} 
                 - \frac{1}{q^0+\omega(\vec q)-i\epsilon} \right),
 \label{eq23}
\end{eqnarray}
with $\omega(\vec q)=\sqrt{\vec q^2+m^2}$.

The $q^0$ integration is done analytically using Cauchy's residues and we finally obtain
\begin{eqnarray}
 &&V_j=-\int\limits_{|\vec{q}|\le q_\mathrm{max}}\frac{d^3q}{(2\pi)^3}\frac{1}{2\omega_j(\vec q)}\frac{1}{2\omega_j(\vec q)}
          \frac{1}{2\omega_\pi(\vec q+\vec k)} \frac{1}{P^0-\omega_j(\vec q)-\omega_j(\vec q)+i\epsilon} 
          \cdot\tilde W_j (2\vec k+\vec q)^2\cdot g^2      \nonumber\\ 
  &&\phantom{ V_j=-\int\limits_{|\vec{q}|\le q_\mathrm{max}} } \cdot
           \left(\frac{1}{P^0-k^0-\omega_j(\vec q)-\omega_\pi(\vec q+\vec k)+i\epsilon} +  
              \frac{1}{k^0-\omega_j(\vec q)-\omega_\pi(\vec q+\vec k)+i\epsilon} \right). 
   \label{eq24}           
\end{eqnarray}
We see that the $d^3q$ integration is constrained by ${|\vec{q}|\le q_\mathrm{max}}$. This is because the $T$ matrix with a 
cutoff $q_{max}$ in the $G$ function implies that the $T$ matrix is of the type 
$T_{q,q^\prime}(\sqrt s)\Theta(q_\mathrm{max}-|\vec q|)\Theta(q_\mathrm{max}-|\vec q^{\,\prime}|)$ with $\vec q$,  
$\vec q^{\,(\prime)}$ the initial and final momenta of the $VV\to VV$ process~\cite{danijuan}. 
We use $ q_\mathrm{max} = 1000  \,\rm MeV$, as found in~\cite{geng}. Since the pion exchanged in the triangle loop of Fig.~\ref{fig2} is off-shell, it is also customary to include a form factor~\cite{Machleidt:1987hj}. The sharp cut-off that we use already eliminates large values of $ \vec{q} $, making moderate the effect of a form factor.

\subsection{Spin considerations}
The vertices of Eq.~(\ref{eq6}) have to be implemented with the spin dependence of the particles. We assume $s$-wave production 
and then the $VV$ produced in $s$-wave and $J=0$ go with the operator ${\vec\epsilon}_{V}{\vec{\epsilon}}_{V^\prime}$
\cite{raquel,geng}. Then the polarization vectors of the $J/\psi$ and $\phi$ must be contracted. Hence, finally the structures
$\langle VVV\rangle$, $\langle VV\rangle \langle V\rangle$ and $\langle V\rangle^3$ of Eq.~(\ref{eq6}) have an extra factor 
\begin{eqnarray}
    {\vec\epsilon}_{J/\psi}{\vec{\epsilon}}_{\phi}{\vec\epsilon}_{V}{\vec{\epsilon}}_{V^\prime}, 
    \label{eq25}
\end{eqnarray}
On the after hand, the $J=0$ projected $T_{ij}$ matrix for $VV^\prime\to V_iV_i^\prime$ has the extra factor~\cite{raquel,geng}
\begin{eqnarray}
    \frac13{\vec\epsilon}_{V}{\vec{\epsilon}}_{V^\prime}{\vec\epsilon}_{V_1}{\vec{\epsilon}}_{V_1^\prime}, 
    \label{eq26}
\end{eqnarray}
This, together with the ${\vec\epsilon}_{V_1}(2\vec k+\vec q)$, ${\vec\epsilon}_{V^\prime_1}(2\vec k+\vec q)$ of the two 
$VPP$ vertices in the triangle loop, when summing over polarizations of the intermediate vectors and considering 
$ \sum_{pl} \epsilon_V^i\epsilon_V^j=\delta_{ij}$, one finds 
\begin{eqnarray}
   {\vec\epsilon}_{J/\psi}{\vec{\epsilon}}_{\phi}{\epsilon}^i_{V}{{\epsilon}}^i_{V^\prime}
    \frac13{\epsilon}^j_{V}{{\epsilon}}^j_{V^\prime}{\epsilon}^l_{V_1}{{\epsilon}}^l_{V_1^\prime}
    {\epsilon}^m_{V_1}{{\epsilon}}^n_{V_1^\prime}(2k+q)^m(2k+q)^n 
    = {\vec\epsilon}_{J/\psi}{\vec{\epsilon}}_{\phi} (2\vec k+\vec q)^2.  
    \label{eq27}
\end{eqnarray}
Thus, finally, the $t_i(M_{\rm inv})$ amplitude of Eq.~(\ref{eq18}) simply has the extra factor 
${\vec\epsilon}_{J/\psi}{\vec{\epsilon}}_{\phi}$. 

The mass distribution for $K^+K^-$, $K^0\bar K^0$ will then be given by  
\begin{eqnarray}
  \frac{d\Gamma_i}{dM_{\rm inv}}=\frac{1}{(2\pi)^3}\frac{1}{4M^2_{J/\psi}} p_\phi\tilde p_{K_i}\overline{\sum}\sum|t_i|^2,
      \label{eq28}
\end{eqnarray}
with 
\begin{align}
    p_{\phi}&=\frac{\lambda^{1/2}(M_{J/\psi}^2,m^2_\phi,M_{\rm inv}^2)}{2M_{J/\psi}},  \label{eq29}    \\
    \tilde{p}_{K_i}&\equiv K_i=\frac{\lambda^{1/2}(M_{\rm inv}^2,m_{K_i}^2,m_{K_i}^2)}{2M_{\rm inv}}. \label{eq30}
\end{align}
In addition, we can have an extra background term for $K^+K^-$, $K^0\bar K^0$ production through direct 
$J/\psi\to \phi K^+K^-(K^0\bar{K}^0)$. This amplitude has $I=0$ in the kaons and is given by 
\begin{eqnarray}
    t_{bac,i}= A {\vec\epsilon}_{J/\psi}{\vec{\epsilon}}_{\phi}; \quad i=K^+K^-, K^0\bar K^0
      \label{eq31}
\end{eqnarray}
which adds coherently to $t_i(M_{\rm inv})$ of Eq.~(\ref{eq18}). Note that, 
$\overline{\sum}\sum{\vec\epsilon}_{J/\psi}{\vec{\epsilon}}_{\phi}=1$, so for practical purpose we can ignore the spins.

\section{Results}   

\begin{figure}
\centering
\includegraphics[width=0.4\columnwidth]{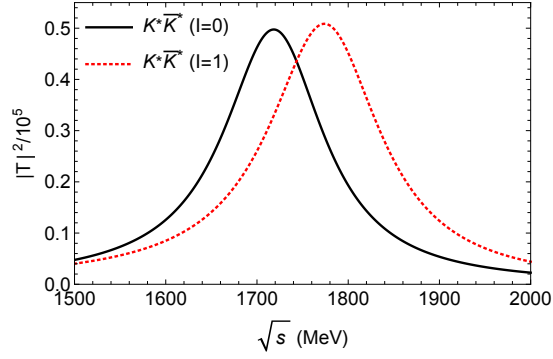}
\caption{Squared modulus of amplitudes $(|T|^2)$ for the elastic reactions of $ K^*\bar{K}^* $ with $I=0$ and $I=1$, according to the Breit-Wigner parametrization of the "real axis" approach reported in Ref.~\cite{geng}.}
\label{tampl}
\end{figure}

In Fig.~\ref{tampl} we show the squared modulus of amplitudes $(|T|^2)$ for $ K^*\bar{K}^* \rightarrow K^*\bar{K}^* $ in $I=0$ and $I=1$ when the average masses of $ K^* $ are used, hence respecting isospin symmetry. We take the masses and widths of Ref.~\cite{geng}, given in Eq.~(\ref{eq11}), and use the Breit-Wigner (BW) parametrization of Eq.~(\ref{eq10}). The couplings $ g_i $ of the resonances to the different channels are taken from Ref.~\cite{geng} and are reproduced in Tables~\ref{table1} and \ref{table2}. The $|T|^2 $ for $I=0$ is about the same as in \cite{geng}, where the amplitudes are taken directly from the unitary coupled-channel approach, $T=[1-VG]^{-1} V$. In the case of $ I=1 $ the BW parametrization produces an amplitude which has about $ 60\% $ higher strength. However, since our conclusions are not qualitatively affected by this difference, we will use this parametrization henceforth.

\begin{figure}
\centering
\includegraphics[width=0.4\columnwidth]{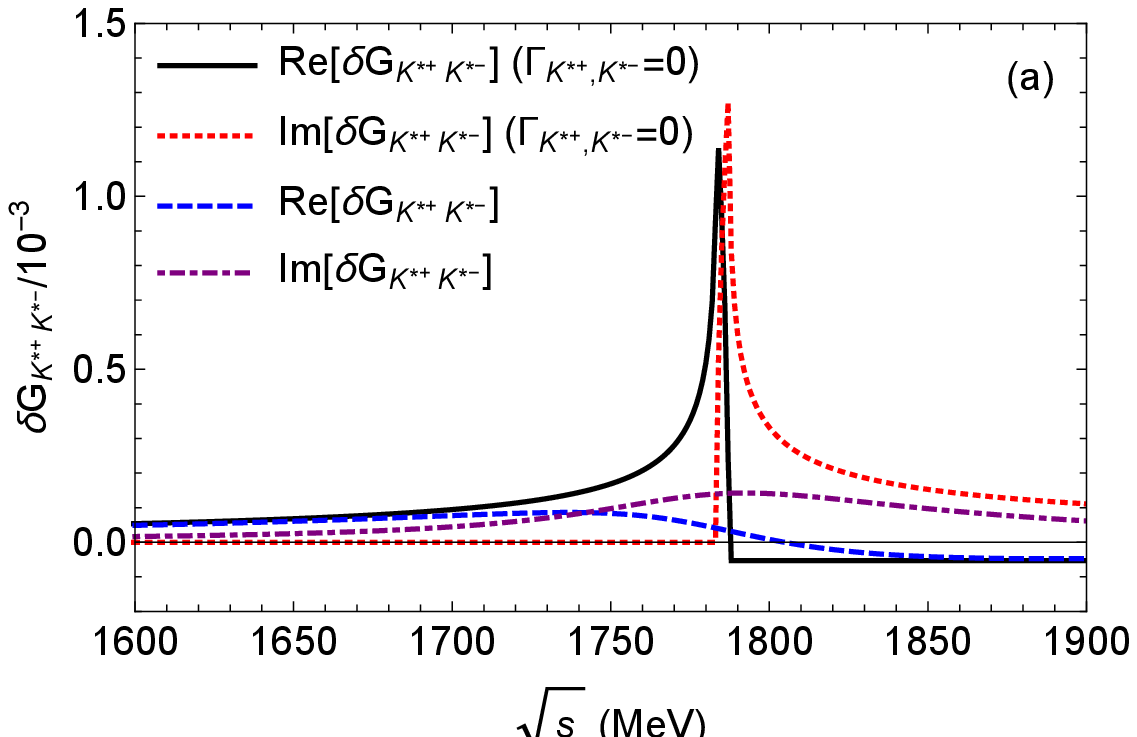}
\includegraphics[width=0.4\columnwidth]{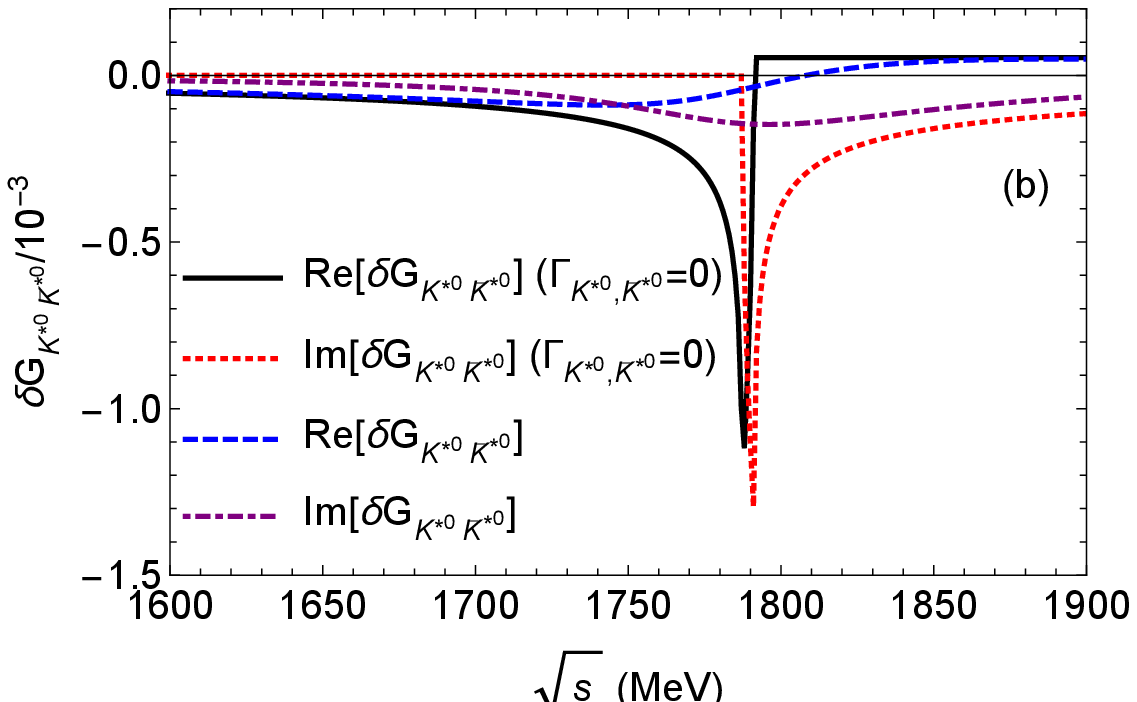}
\caption{Real and imaginary parts of $ \delta G $ for the channels $  K^{*+} K^{*-} (a)$  and  $ K^{*0}\bar{K}^{*0} (b)$.}
\label{ggtilde}
\end{figure}

In Fig.~\ref{ggtilde} we plot the differences $ \delta G_{K^{*+} K^{*-}} $ and $ \delta G_{K^{*0}\bar{K}^{*0}} $ as defined in 
Eq.~(\ref{eq17}). The calculations, done ignoring the widths of the vector kaons and antikaons, correspond to the sharp structures, while considering widths they are the smooth ones (in Appendix~\ref{AppA} we describe the procedure to implement the width in the loop functions). We observe that the  $ \delta G $'s are small but this must be gauged versus G, which is of the order of $ 10^{-3} $, giving changes by about a factor of $ 10-20 \% $. Besides, the differences have opposite sign for  $ K^{*0}\bar{K}^{*0} $ and $ K^{*+} K^{*-}$.

\begin{figure}
\centering
\includegraphics[width=0.4\columnwidth]{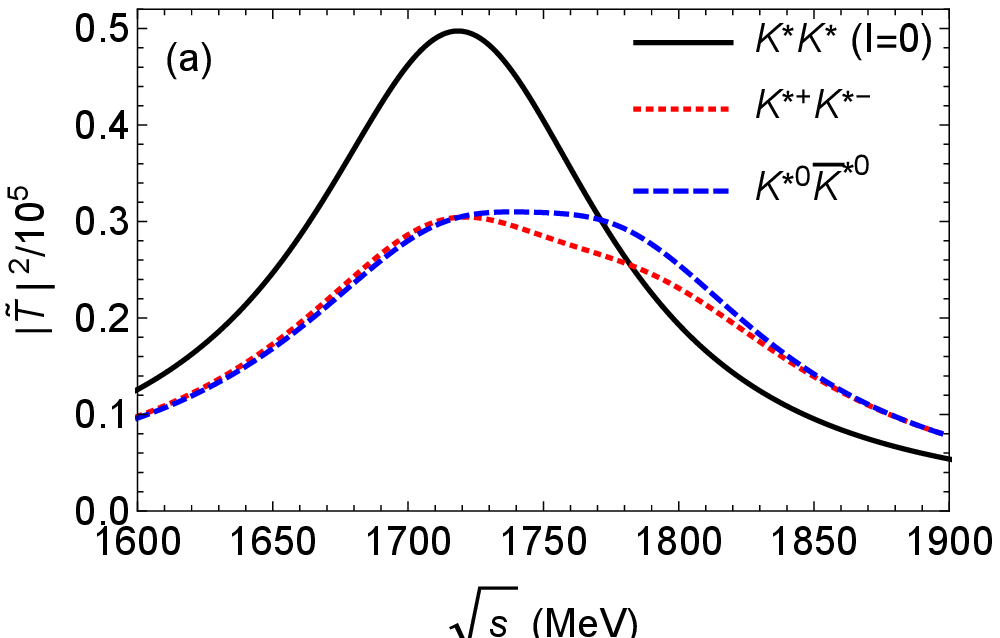}
\includegraphics[width=0.4\columnwidth]{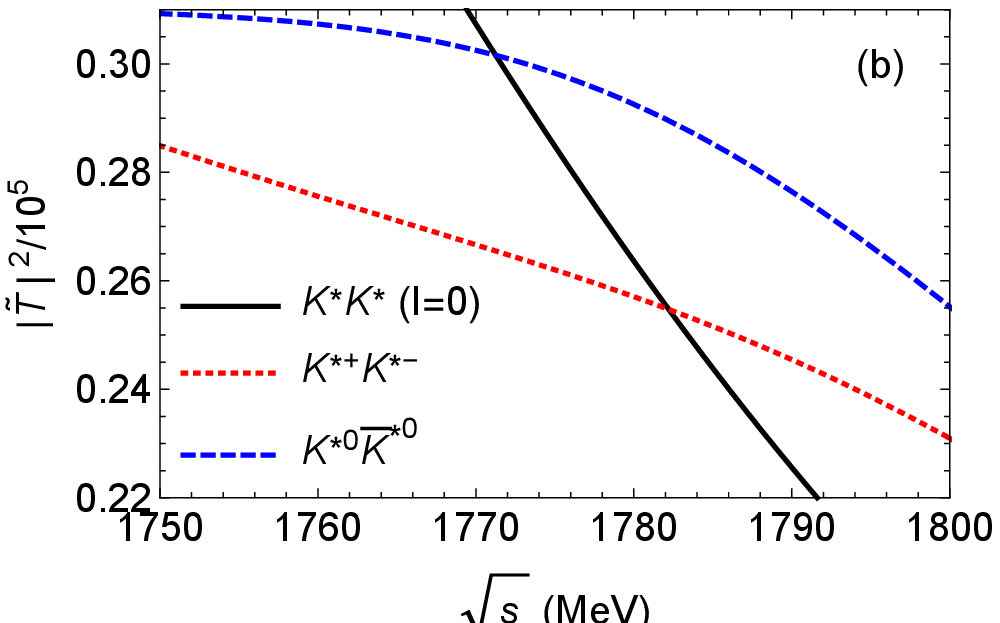}
\caption{(a) Squared  modulus of $ \tilde{T} $ amplitudes for the channels $ K^{*+} K^{*-}$ and $ K^{*0}\bar{K}^{*0} $. (b) Same as in (a) but with a zoom of a narrow energy window.}
\label{ttildeampl}
\end{figure}

In Fig.~\ref{ttildeampl} we show the squared modulus of the amplitudes $(| \tilde{T} |^2)$ after the isospin breaking is considered  for the elastic channels $ K^{*+} K^{*-}$ and $ K^{*0}\bar{K}^{*0} $, taking into account the widths of these kaon vectors and of the $\rho $ meson (the widths of the $\omega$ and $ \phi $ mesons are neglected due to their smallness). We see indeed that there are changes in the shape of the $ K^{*+} K^{*-}$ and $ K^{*0}\bar{K}^{*0} $ amplitudes, because of the isospin mixing. Should we have pure $I=0$ or $I=1$, both amplitudes would be identical. We also show there $|T|^2$ for the $I=0$ case for comparison.

\begin{figure}
\centering
\includegraphics[width=0.4\columnwidth]{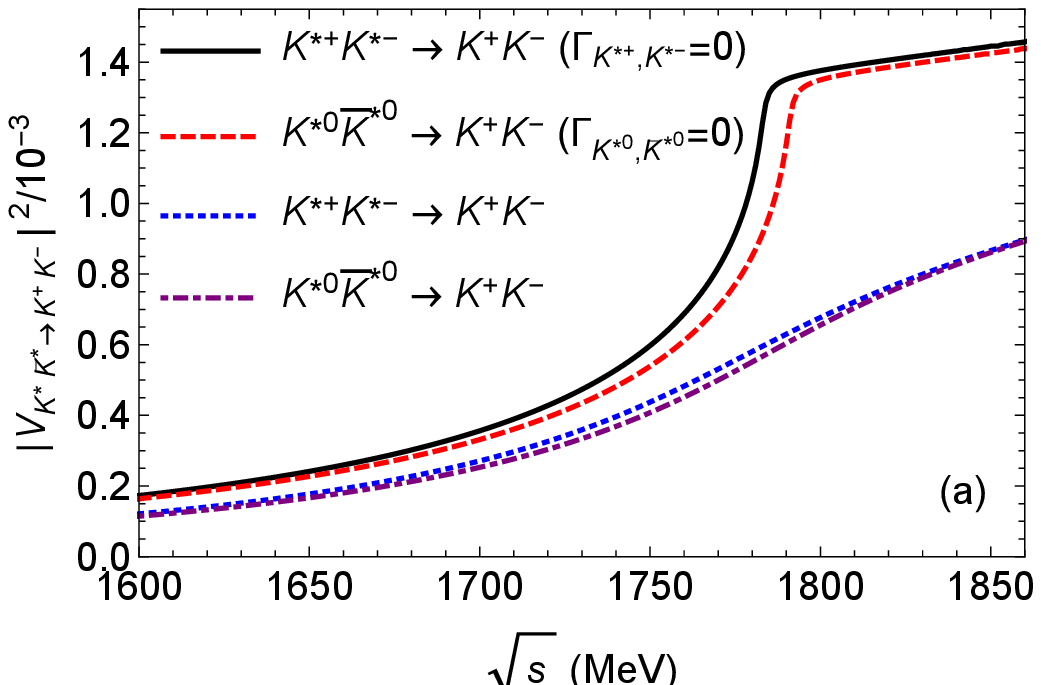}
\includegraphics[width=0.4\columnwidth]{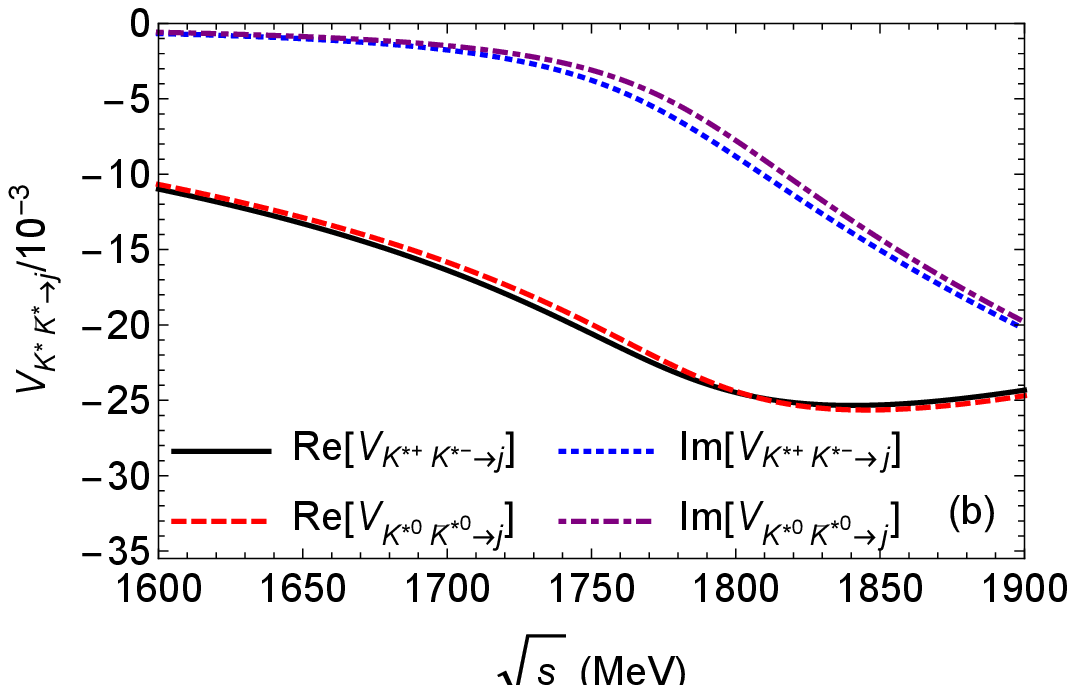} 
\caption{(a) Squared  modulus of triangle loop integrals $ V_{K^{*+} K^{*-}, j } $ and $ V_{K^{*0}\bar{K}^{*0} , j } $, with the final state $ j= K^{+} K^{-}$.  (b) Real and imaginary parts of  $ V_{K^{*+} K^{*-}, j } $ and $ V_{K^{*0}\bar{K}^{*0} , j } $, including the widths of the vector kaons. We have factorized the factor $ g^2 $ present in Eq.~(\ref{eq24}).}
\label{triangleloop}
\end{figure}

Fig.~\ref{triangleloop} presents the results for the triangle loops $ V_{K^{*+} K^{*-}, j } $ and $ V_{K^{*0}\bar{K}^{*0} \rightarrow j } $; since the cases $ j= K^{+} K^{-}, K^{0}\bar{K}^{0} $ engender practically identical outcomes, we only show those for $ j= K^{+} K^{-} $. The squared modulus of the triangle loop integrals is displayed in Fig.~\ref{triangleloop} (a).
The two upper (lower) curves are obtained with vanishing (nonzero) widths of the kaon vectors. We observe that the inclusion of the widths smooths the curve and the differences become small. For a more detailed look of this behavior, in Fig.~\ref{triangleloop} (b) the real and imaginary parts of the triangle loop are shown separately considering the $ K^{*} $ widths.

\begin{figure}
\centering
\includegraphics[width=0.4\columnwidth]{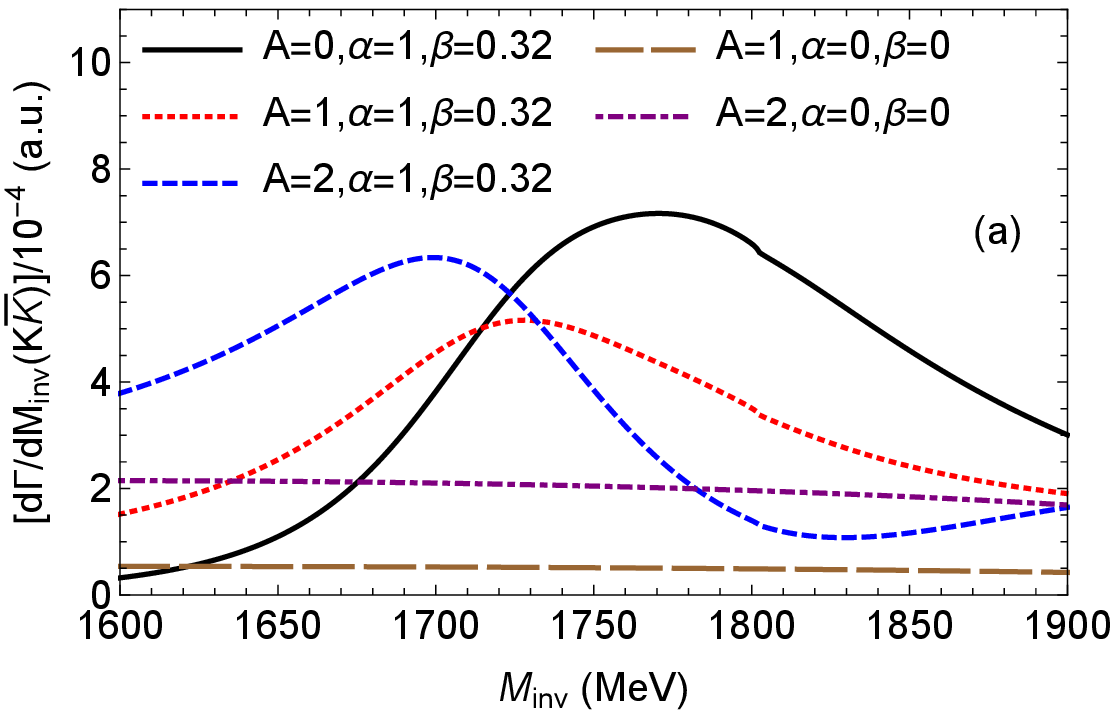}
\includegraphics[width=0.4\columnwidth]{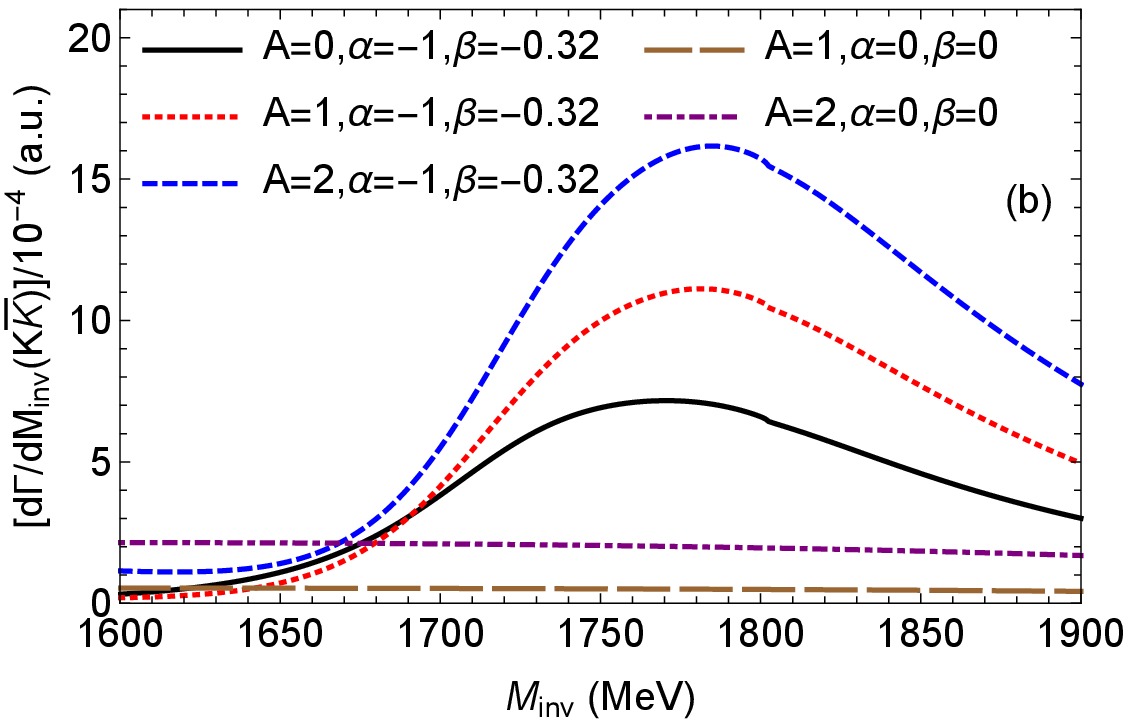}
\caption{Differential decays for the reaction $J/\psi \rightarrow \phi K^+ K^- $, taking distinct parametrizations for the parameter $A$. The curves representing the background encoded only in the parameter $ A $ (i.e. $ \alpha, \beta = 0 $) have been also shown. }
\label{diffdecays1}
\end{figure}

\begin{figure}
\centering
\includegraphics[width=0.4\columnwidth]{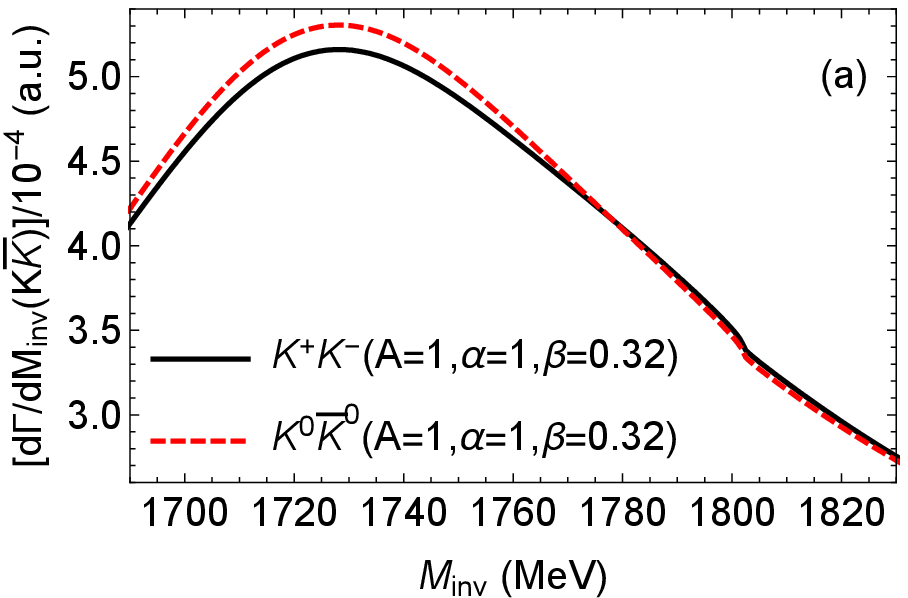}
\includegraphics[width=0.4\columnwidth]{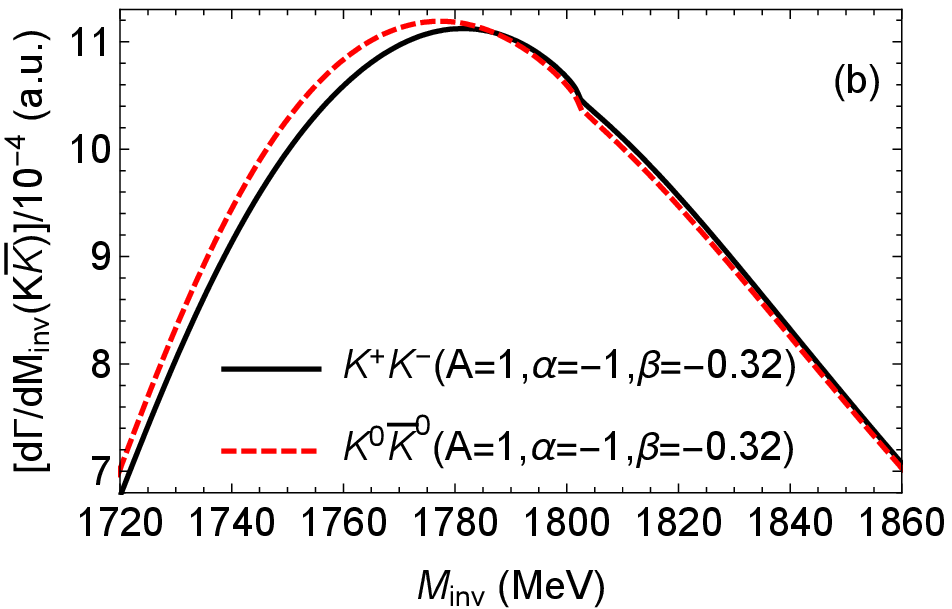} \\
\caption{Differential decays for the reactions $J/\psi \rightarrow \phi K^+ K^- (K^0 \bar{K}^0) $, taking distinct parametrizations. }
\label{diffdecays2}
\end{figure}

We plot in Fig.~\ref{diffdecays1} the mass distributions $ d \Gamma / d M_{inv} $ for  $ K^{+} K^{-}$ production of 
Eq.~(\ref{eq28}) with distinct parametrizations ($ \gamma =0 $ in all cases): $(i) A=0, \alpha =1, \beta =0.32 $; $(ii) A=1, \alpha =1, \beta =0.32 $; $(iii) A=2, \alpha =1, \beta =0.32 $; $(iv) A=0, \alpha =-1, \beta =-0.32 $ (identical to the case $ (i) $);  $(v) A=1, \alpha =-1, \beta =-0.32 $ and $(vi) A=2, \alpha =-1, \beta =-0.32 $. In Fig.~\ref{diffdecays2} the $ d \Gamma / d M_{inv} $ is displayed in a smaller energy window for both $ K^{0}\bar{K}^{0} $ and $ K^{+} K^{-}$ production. We remark that the values of the parameters $\alpha, \beta, \gamma$ have been chosen by benefiting from the findings reported in Ref.~\cite{raqueldai}, where the $ J/\psi $ decay into $\phi (\omega)$ plus vector-vector molecular states  have been analyzed. The value of the background encoded in the parameter $ A $ has also been chosen such that for $ A=2 $ alone (i.e. $ \alpha, \beta = 0 $) one has a strength similar to that of $ \alpha = 1, \beta = 0.32 $ alone (i.e. $ A=0$).

\begin{figure}
\centering
\includegraphics[width=0.4\columnwidth]{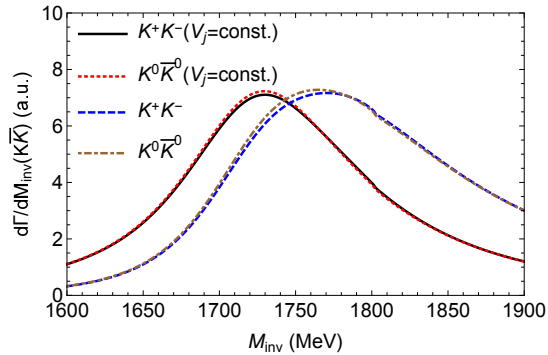}
\caption{Differential decays for the reactions $J/\psi \rightarrow \phi K^+ K^- , K^0 \bar{K}^0  $ obtained with and without considering the triangle loop contributions. We have used the parametrization  $(iv) A=0, \alpha =-1, \beta =-0.32 $.  In the case without the triangle contribution, we have replaced $ V_{K^{*} \bar{K}^{*}, j } $ by a factor ($-2.0 \times 10^{-2}$) to have a similar magnitude to the situation with the triangle loop. }
\label{diffdecays3}
\end{figure}

In order to have a better understanding of the influence of the triangle loop integral on the results reported above, in Fig.~\ref{diffdecays3} we compare the differential decays for the reactions $J/\psi \rightarrow \phi K^+ K^- , K^0 \bar{K}^0 $ obtained considering the triangle loops with those without the triangle contributions. It can be seen clearly that without the triangle loop the peak near the mass of the $ I=1 $ resonance disappears, which tells  us that the mixing between the $ I=0 $ and $ I=1 $ contribution might not play a relevant role as a possible source of isospin breaking. The peak around $1780$ MeV of the mass distribution is caused by 
the triangle loop function and not by the isospin mixing.

\begin{figure}
\centering
\includegraphics[width=0.4\columnwidth]{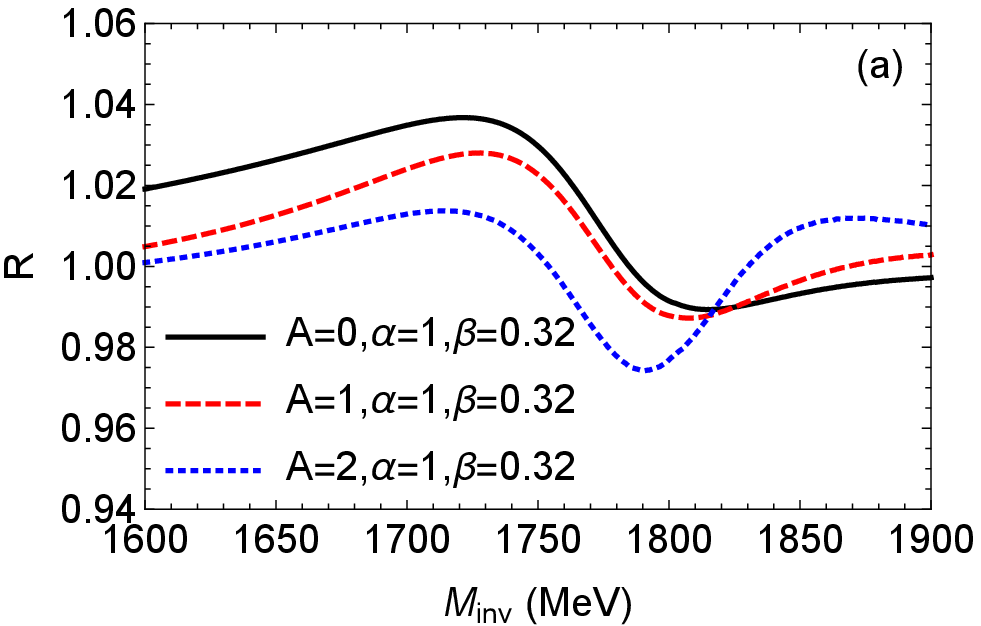}
\includegraphics[width=0.4\columnwidth]{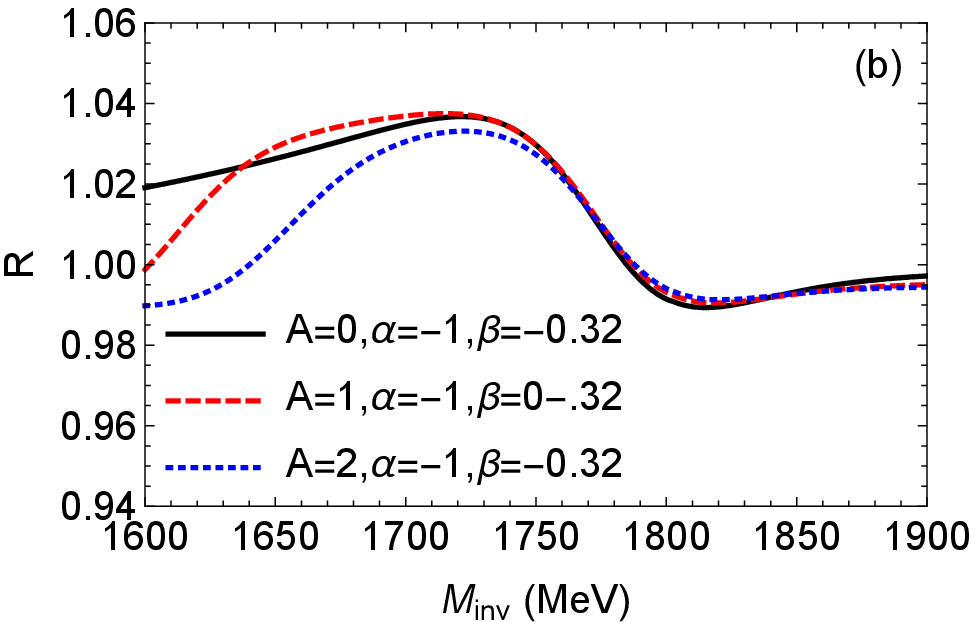}
\caption{Ratio between the differential decays for the reactions $J/\psi \rightarrow \phi K^+ K^- $ and $J/\psi \rightarrow \phi  K^0 \bar{K}^0 $, taking distinct parametrizations.}
\label{ratio}
\end{figure}

Finally, in Fig.~\ref{ratio} we show the ratio of the $K^+K^-$ and $K^0\bar K^0$ production rates. They show deviations of about
$5\%$ from unity, but there is a distinct behavior, with the differences changing sign around $1780$ MeV, which are due to the isospin
mixing and which could serve to show the position of the $a_0$ resonance.
The calculation of the vertices $V_i$ and the mass distributions are done using the sharp cut-off. We have seen the effect of introducing an extra off-shell pion form factor in momentum space $ \Lambda^2 /(\Lambda^2 + \vec{q}\,^2) $ in Eq.~(\ref{eq24}) with $\Lambda = 1250 \, \rm MeV$, and we find a reduction of about $30\%$ in the strength of  $ |V_i| ^2 $ and  $d\Gamma_i / dM_{\rm inv}$, affecting equally the 
$ K^+ K^-  $ and $K^0 \bar{K}^0 $ production, which is not relevant since we make the plots in arbitrary units. The ratios of Fig.~\ref{ratio}, a main conclusion of the paper, remain practically unchanged. 

\section{Conclusions}
\label{sec-con}
We have addressed the issue of a possible isospin violation in the strong decay $J/\psi \to \phi K \bar K$, looking for differences in the production rates of $K^+ K^-$ or $K^0 \bar K^0$ in the region of $1700$-$1800$ MeV. The idea is that in this region there are two resonances coupling to  $K \bar K$, the $f_0(1710)$ and another new $a_0$ resonance, seen in \babar and BESIII experiments in the 1700-1800 MeV region. In dynamical models of generation of resonances from interaction of mesons, these two resonances are generated from the interaction of mostly $K^* \bar K^*$ together with other coupled channels. With $J/\psi$ and  $\phi$ having isospin zero, the $K^+ K^-$ or $K^0 \bar K^0$ should also be produced in $I=0$ and hence we should see equal rates for $K^+ K^-$ or $K^0 \bar K^0$ production. The idea of an isospin violation stems from the realization that the charged and neutral $K^*$ have different masses. This has as a consequence that the $K^* \bar K^*$  loops present in the construction of the $K^* \bar K^*$ scattering matrix, and in the reaction mechanism leading to the $J/\psi \to \phi K \bar K$ decay, are different and hence we can expect a final different rate in the production of $K^+ K^-$ or $K^0 \bar K^0$. While some differences are seen if we take the nominal masses of the $K^*$ states, the difference of these masses are small compared to the widths of the $K^*$ particles, and as soon as the latter are considered in the formalism, the production rates of of $K^+ K^-$ or $K^0 \bar K^0$ are practically equal, with differences below 5\%. 
  
   On the other hand we find interesting shapes in the $K \bar K$ mass distributions which one must be careful to interpret. In order to produce the $K^+ K^-$ or $K^0 \bar K^0$ through resonances, first we have to create a $\phi$ and a couple of vectors which can couple to the resonances. These mesons then interact and produce the resonance through a $ VV \to VV$ collision. Finally the resulting VV pair must convert into a $K \bar K$ pair. This is done through a triangle loop in which the two VV propagate and exchange a pseudoscalar to finally convert into the $K \bar K$ pair. The exchanged pseudoscalar in the triangle loop is far off shell, to the point that if one factorized it outside the triangle integral (we do not do this approximation) the remaining integral is nothing but the loop G function of $K^* \bar K^*$, which has a singularity at threshold, and $|G|^2$ peaks at that threshold. Smoothed by the effects of the $K^*$ width, the peak softens but has as a consequence to produce an enhancement close to the  $K^* \bar K^*$ threshold which happens to be around the position of the $I=1$ resonance. This shape, however, can change if there is a sizeable amount of direct, non resonant $J/\psi \to \phi K \bar K$, without intermediate VV production. Upon interference with the resonant mechanism this can produce different shapes in the  $K \bar K$ mass distributions, yet, without changing the relative rates of  $K^+ K^-$ or 
$K^0 \bar K^0$ production which are about the same.  

   In summary, we can conclude that the shapes of the $K \bar K$ mass distributions can teach us much about the reaction mechanism, which is interesting in itself, but we should not expect much difference in the rates of $K^+ K^-$ or $K^0 \bar K^0$ production since the isospin violation found is minimal. 

\begin{acknowledgments}
We appreciate discussions and encouragement to do this work from Nils H\"{u}sken. LMA has received funding from the Brazilian agencies Conselho Nacional
de Desenvolvimento Cient\'ifico e Tecnol\'ogico (CNPq) under contracts 309950/2020-1, 400215/2022-5, 200567/2022-5), 
and Funda\c{c}\~ao de Amparo \'a Pesquisa do Estado da Bahia (FAPESB) under the contract INT0007/2016.
This work was supported in part by the National Natural Science Foundation of China under Grant No. 12147215. 
This work was partly supported by the Spanish Ministerio de Economia y Competitividad (MINECO) and 16 European FEDER
funds under Contracts No. PID2020-112777GB-I00, and by Generalitat Valenciana under Contract No. PROMETEO/2020/023. 
This project has received funding from the European Union Horizon 2020 research and innovation programme under the program H2020-INFRAIA-2018-1, Grant Agreement No. 824093 of the STRONG-2020 project.
\end{acknowledgments}

\appendix
\section{$G$ function and triangle loop with $ K^{*} $ widths}
\label{AppA}

\subsection{$G$ function}

\begin{figure}
\centering
\includegraphics[width=5cm]{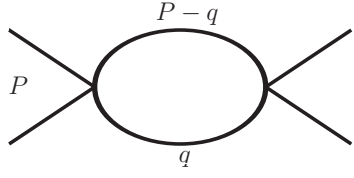}
\caption{Diagram corresponting to the loop function for $ K^{*}\bar{K}^{*} $.}
\label{figgfunction}
\end{figure}

Here we consider the $G$ function of $ K^{*}\bar{K}^{*} $, which is the loop function represented in Fig.~\ref{figgfunction}. Its expression is given by
\begin{eqnarray}
G (\sqrt{s}) &= & i \int \frac{d^4 q}{(2\pi)^4} \frac{1}{\left(q^2-m_{K_1^{*}}^2+i\epsilon\right)\left[(P-q)^2-m_{K_2^{*}}^2+i \epsilon \right]}, \nonumber \\
\label{eqgfunction1}
\end{eqnarray}
where $ P = (P^0 , \vec{P})= (\sqrt{s} , \vec{0})$. We split the propagators into their positive and negative energy parts, as done in 
Eq.~(\ref{eq23}), and keep their positive-energy part since the vector kaons are massive particles. This allows us to rewrite 
Eq.~(\ref{eqgfunction1}) as 
\begin{eqnarray}
G (\sqrt{s}) &= & i \int \frac{d^4 q}{(2\pi)^4} 
\frac{1}{2 \omega_1(q)  2 \omega_2(q)} 
\frac{1}{\left[q^0-\omega_1(\vec{q})+i \epsilon \right]\left[P^0-q^0 -\omega_2(\vec{q})+i \epsilon  \right]} ,
\label{eqgfunction2}
\end{eqnarray}
with $ \omega_{i=1,2} (\vec{q}) \equiv \sqrt{\vec{q}^2 + m_{K_i^{*}} ^2 }$. Then, integrating analytically over $ q^0 $ via the Cauchy's residue theorem with the contour chosen in order to pick up the residues at $ q^0 = \omega_{i} (\vec{q}) $ (which places this particle on-shell), we obtain
\begin{eqnarray}
G (\sqrt{s}) &= &  \int \frac{d^3 q}{(2\pi)^3} 
\frac{1}{2 \omega_1(\vec{q})  2 \omega_2(\vec{q})} 
\frac{1}{P^0 - \omega_1(\vec{q}) -  \omega_2(\vec{q}) +i \epsilon } .
\label{eqgfunction3}
\end{eqnarray}

The widths of the particles are included in this approach by replacing the propagators as follows:
\begin{eqnarray}
\frac{1}{q^0-\omega_1(\vec{q})+i \epsilon} & \rightarrow & \frac{1}{q^0 -\omega_1(\vec{q}) + i \frac{\Gamma (s_1)}{2}} , \nonumber \\
\frac{1}{P^0-q^0 - \omega_2(\vec{q}) + i \epsilon} & \rightarrow & \frac{1}{P^0 - q^0 - \omega_2(\vec{q})+i \frac{\Gamma (s_2)}{2}};
\label{rep-prop}
\end{eqnarray}
and taking into account that the particle 1 has momentum $ q $ on-shell in the Cauchy integration, the variables $ s_1 $ and $ s_2 $ can be written as  
\begin{eqnarray}
s_1 & = & (q^0)^2 - \vec{q}^2 \rightarrow \omega_1(\vec{q})^2 - \vec{q}^2 = m_{K_1^{*}}^2, \nonumber \\
s_2 & = & (P^0 - q^0)^2 - \vec{q}^2 \rightarrow (P^0 - \omega_1(\vec{q}))^2 - \vec{q}^2
\nonumber \\
& = & s + m_{K_1^{*}} ^2 - 2  \omega_1(\vec{q}) \sqrt{s}. 
\label{s1s2}
\end{eqnarray}
In this sense, the $\Gamma (s_i)$ function is given by (see Ref.~\cite{geng} for a more detailed discussion)
\begin{eqnarray}
\Gamma (s_i) & = & \Gamma(m_{K_i^{*}}^2)\left( \frac{m_{K_i^{*}}^2}{s_i} \right) \left( \frac{p(s_i)}{p(m_{K_i^{*}}^2)} \right)^3
\Theta ( \sqrt{s_i} - m_1 -m_2), 
\label{gammasi}
\end{eqnarray}
where 
\begin{eqnarray}
p(s_i) & = & \frac{\lambda ^{1/2}\left(s_i, m_1^2,m_2^2\right)}{2 \sqrt{s_i}}, \nonumber \\
p\left(m_{K_i^{*}}^2\right) & = & \frac{\lambda ^{1/2}\left( m_{K_i^{*}}^2, m_1^2,m_2^2\right)}{2 m_{K_i^{*}}}, 
\label{psipmks}
\end{eqnarray}
with  $ m_1= m_{K_i} $ and  $ m_2= m_{\pi} $ for particles in the loop being vector kaons, and $ \lambda  $ is the K\"allen function. 

Hence, with all ingredients described above, Eq.~(\ref{eqgfunction3}) assumes the final form  
\begin{eqnarray}
G (\sqrt{s}) &= & \int \frac{d^3 q}{(2\pi)^3} 
\frac{1}{2 \omega_1(\vec{q})  2 \omega_2(\vec{q})} 
\frac{1}{\sqrt{s} - \omega_1(\vec{q}) -  \omega_2(\vec{q}) + i \frac{\Gamma (s_1)}{2} +i \frac{\Gamma (s_2)}{2}  } .
\label{eqgfunction4}
\end{eqnarray}

\subsection{Triangle loop integral}

We follow the same approach as in the former subsection. Accordingly, we simply replace 
in Eq.~(\ref{eq24})
\begin{eqnarray}
\frac{1}{P^0 - \omega_1(\vec{q}) -  \omega_2(\vec{q}) + i \epsilon  } \rightarrow \frac{1}{P^0 - \omega_1(\vec{q}) -  \omega_2(\vec{q}) + i \frac{\Gamma (s_1)}{2} +i \frac{\Gamma (s_2)}{2}  } .
\label{eqtrianglefunction1}
\end{eqnarray}


\end{document}